\documentstyle[12pt,aaspp4,epsf]{article}

\lefthead{Wandel, Peterson, and Malkan}
\righthead{Central Masses of AGN}

\def\mfig #1#2#3#4{\par
\epsfxsize=#1 cm
\moveright #2cm
\vbox{\epsfbox{#3}}
{\noindent Figure~#4 }\vskip .3cm }

\def\myfig #1#2#3#4{
\vskip #1 cm
\epsfxsize=#1 cm
\moveright #2cm
\vbox{\epsfbox{#3}}
{\noindent Figure~#4 }\vskip .3cm 
}
\def\lg{{\rm log}}
\def\bh{black hole}

\def\el {emission-line}
\def\bhs{black~holes}

\def\ed {Eddington~}
\def\ers{{\rm erg/sec}}
\def\ms{M_{\odot}}
\def\et{et al.\ }

\def\rev{reverberation~}
\def\vFWHM{\ifmmode v_{\mbox{\tiny FWHM}} \else
            $v_{\mbox{\tiny FWHM}}$\fi}
\def\kms{\ifmmode {\rm km\ s}^{-1} \else km s$^{-1}$\fi}
\def\ers{\ifmmode {\rm erg\ s}^{-1} \else erg s$^{-1}$\fi}

\begin{document}
\hfill updated \today

\vskip 0.1in

\title{The Black Hole to Bulge Mass Relation in Active Galactic Nuclei 
}
\author{A. Wandel\altaffilmark{1}}
\affil{Racah Institute of Physics, The Hebrew University, Jerusalem 91904, Israel}

\altaffiltext{1}{On leave at the Department od Physics and Astronomy, University of California, Los Angeles, CA 90095-1562; E-mail: wandel@astro.ucla.edu}

\begin {abstract}

The masses of the central black holes in Active Galactic Nuclei (AGNs) can be estimated 
using the broad emission-lines as a probe of the virial mass inside the BLR. 
Using reverberation mapping to determine the size of the Broad Line Region (BLR) 
and the width of the variable component of the line profile  H$\beta$ line 
it is possible to find quite accurate virial mass estimates for AGN with adequate data. 
Compiling a sample of AGN with reliable central masses and bulge magnitudes 
we find an average black hole to bulge mass ratio 
of 0.0003, a factor of 20 less than the value found for normal
galaxies and for bright quasars.
This lower ratio is more consistent with the black hole mass density predicted
from quasar light, and agrees with the value found for our Galaxy. We argue that the black hole to bulge mass ratio
actually has a significantly larger range than indicated by MBHs detected 
 in normal galaxies (using stellar dynamics) and in bright quasars, 
which may be biased towards larger black holes , 
and derive a scenario of black hole growth that explains the observed distribution.

\end {abstract}


\section{Introduction}
Massive \bh s (MBHs) have been postulated in quasars and active galaxies
(Lynden-Bell 1969, Rees 1984).
Evidence for the existence of MBHs has recently been found in the center of
our Galaxy (Ghez \et 1998, Genzel \et 1997) and in the weakly active galaxy
NGC 4258 (Miyoshi \et 1995).
Compact dark masses, probably MBHs, have been detected in the cores of 
many normal galaxies using stellar dynamics (Kormendy and Richstone 1995).
The MBH mass appears to correlate with the galactic
bulge luminosity, with the MBH 
 being about one percent of the mass of the spheroidal bulge
(Magorrian \et 1998, Richstone \et 1998). 

The question whether AGN follow a similar \bh -bulge relation  
as normal galaxies is a very interesting one, as it may shed light on the connection between the host galaxy and the active nucleus.
Wandel \& Mushotzky (1986) have found an excellent correlation between the 
virial mass included within the narrow line region (of order of tens to hundreds
pc from the center) and the \bh~mass estimated from X-ray variability in a sample of Seyfert 1 galaxies.
A \bh -bulge relation similar to that of normal galaxies 
has been reported between MBH of bright quasars and the 
bulge of their host galaxies (Laor 1998), but the \bh~ and 
bulge mass estimates have large uncertainties (section 3.2). 

Seyfert 1 galaxies provide an opportunity to obtain more reliable 
\bh -to- bulge mass ratios (BBRs): because of their lower nuclear brightness,
their bulge magnitudes can be measured directly. Also the \bh~ mass estimates
(Wandel 1998) are much more reliable for AGN with reverberation
data, which are more readily obtained for low luminosity AGN.
The relation between the bulge and the nonstellar central source has been  
studied for many Seyfert 
galaxies (Whittle 1992; Nelson \& Whittle 1996). These works find 
a tight correlation between the stellar velocity dispersion and the
O[III] line and radio luminosity.

Reliable BLR size measurements are now possible through reverberation
mapping techniques (Blandford \& McKee 1982, recently reviewed by Netzer \& Peterson 1997).
High quality reverberation data and virial masses are presently available 
 for about twenty AGN, most of them Seyefert 1s (Wandel, Peterson and Malkan 1999, hereafter WPM). We combine the \rev masses (section 2) with Whittle's bulge estimates
in order to study the BBR in low-luminosity AGN and compare it to MBHs in
normal galaxies and quasars (section 3). In section 4 we derive a 
MBH-evolution theory that can explain our results.

\section{BLR reverberation as a probe of \bh masses in AGN}

Broad emission lines probably provide the
best probe of \bh s in AGN. 
Assuming the line-emitting matter is gravitationally bound, 
and hence has a near-Keplerian velocity dispersion (indicated by the line 
width), it is possible to estimate the virial central mass:
$M\approx G^{-1}rv^2 .$
 This remains true
 for many models where the line emitting gas is not in Keplerian motion,
such as radiation-driven motions and disk-wind models (e.g. Murray \et 1998):
in a diverging outflow the density (end hence the emissivity) decreases
outwards, the emission is dominated by the gas 
close to the base of the flow, where the velocity is close to the escape 
velocity. 
(Note that if the velocity is actually larger
than Keplerian, the virial mass is an upper limit and the result that the Seyfert galaxies in our sample 
have smaller \bh~ masses than MBHs detected in 
normal galaxies becomes even stronger).

The main problem in estimating the virial mass from the \el data is to obtain
a reliable estimate of the size of the BLR, and to correctly identify the
line width with the velocity dispersion in the gas. 
WPM use the continuum/emission-line cross-correlation function to measure the
responsivity-weighted radius $c\tau$ of the BLR (Koratkar \& Gaskell 1991), 
and the variable (rms) component of the spectrum to measure the velocity dispersion in the same
part of the gas which is used to calculate the BLR size,
 automatically 
excluding constant features such as narrow emission lines and 
Galactic absorption.
The line width and the BLR size yield the virial
"reverberation" mass estimate 
$M_{rev} \approx (1.45\times 10^5\ms )  c\tau_{days}v_3^2$  
where
$v_3$ is the rms FWHM in units of $10^3 \kms$.

The virial assumption ($v \propto r^{-1/2}$) has been directly tested using
 data for NGC 5548 (Krolik et al.\ 1991; 
Peterson \& Wandel 1999). 
The latter authors find that when the BLR reverberation size
is combined with the rms line width in multi-year data for NGC 5548, the
virial masses derived from different emission lines and epochs are 
all consistent with a single value ($(6.3\pm 2)\times 10^7\ms$)
which demonstrates the case for a
Keplerian velocity dispersion in the line-width/time-delay data.

\section {The Black-Hole - Bulge Relation }
\subsection{Seyfert 1 galaxies}

We use the WPM sample with the virial mass derived from the H$\beta$ line by the reverberation-rms method (table 1). 
For 13 of the objects in the WPM sample we obtain the bulge magnitudes
from the compilation of Whittle \et (1992), who calculate
the bulge magnitude from the total blue magnitude, using 
the empirical formula of Simien \& deVaucolours (1986), relating the galaxy
type to the bulge/total fraction. The bulge magnitudes are corrected for the nonstellar emission using the correlation between H$\beta$ and the 
nonstellar continuum luminosity (Shuder 1981).  

Mkn 110 and Mkn 335 have no estimated bulge magnitude in Whittle's compilation, 
because they do not have well defined Hubble types. For these objects
we adopt a canonical Hubble type of Sa, which
has a bulge correction ($m_{bulge}-m_{gal}$) of 1.02 mag. For Mkn 335 there is already a fairly
large (and therefore uncertain) correction for the active nucleus (1.17 mags).
For 3C120 the bulge magnitude is taken from Nelson \& Whittle (1995) who
find a bulge magnitude of -22.12.
The uncertainties in the bulge magnitude were estimated from Whittle's (1992)
quality indicators. These indicators estimate the error in the subtraction
of the nonstellar luminosity and some other factors, which for most galaxies
amount to an uncertainty in the range 0.2-0.6 magnitudes. To the galaxies with
an uncertain Hubble type we assign an uncertainty of 1.2 mag. 

\begin{table}
\caption{AGN Central Masses derived from BLR data, Compared with 
Host Bulge magnitudes and corresponding mass.
Column (2) -- FWHM of (H$\beta$), rms profile, in $10^3$km/s. (3) -
 log(lag) -- corresponding to the BLR size light days,
(4) -- absolute blue bulge magnitude from Whittle \et (1992),
(5) -- log of the galactic bulge mass ($M_{bul}$) in $\ms$, 
(6) -- \bh mass (from WPM), 
(7) -- BH to bulge mass ratio, (8) -- the Eddinton ratio 
of the ionizing luminosity.
 \label{tbl-2}}
\begin{center}
\begin{tabular}{llllllll}
\tableline
{}&{}&{}&{}&{}&{}&{}&{}\\
Name   &  FWHM &   log($\tau$)& $-M_B$
&  log$(M_{bul})$ & ${M_{bh}\over 10^7\ms}$ 
&log(${M_{bh}\over M_{bul}}$)& log(${L_{ion}\over L_E}$)\\
{(1)}&{(2)}&{(3)}&{(4)}&{(5)}&{(6)}&{(7)}&(8)\\
\tableline
{}&{}&{}&{}&{}&{}&{}&{}\\

3C\,120\tablenotemark{a}     &$  2.2$& 1.64& 20.3$\pm$ 1.2  & 10.77$\pm$ 0.5  &  $ 3.1^{+2.0}_{-1.5}$ &-3.28&-0.57\\
3C\,390.3   &$  10.5$& 1.38& 22.12$\pm$0.4 &  11.0$\pm$ 0.15  &  $39.1^{+12.4}_{-14.8}$ &-2.41&-2.19\\
Akn\,120    &$  5.85$& 1.59& 21.06$\pm$0.8& 11.12$\pm$0.3  &  $19.4^{+4.1}_{-4.6}$&-2.83 &-1.48\\
F\,9        &$  5.9$& 1.23& 22.25$\pm$0.6& 11.67$\pm$0.25  &  $ 8.7^{+2.6}_{-4.5}$ &-3.73&-1.84\\
IC\,4329A   &$ 5.96$& 0.15& 19.93$\pm$0.8& 10.60$\pm$0.3  &  $<0.73$ &$<-3.73$ &$>-2.94$\\
Mrk\,79     &$  6.28$& 1.26& 20.19$\pm$0.2& 10.72$\pm$0.1  &  $10.4^{+4.0}_{-5.7}$ &-2.70&-1.87\\
Mrk\,110\tablenotemark{a}    &$  1.67$& 1.29& 20.76$\pm$1.0& 10.98$\pm$0.4 &  $0.80^{+0.29}_{-0.30}$ &-4.07&-0.69 \\
Mrk\,335\tablenotemark{a}    &$  1.26$& 1.23&  20.02$\pm$1.0 & 10.64$\pm$0.4& $0.39^{+0.14}_{-0.11}$ &-4.06&-0.49\\
Mrk\,509    &$  2.86$& 1.90& 21.75 $\pm$0.6& 11.44$\pm$0.25  &  $9.5^{+1.1}_{-1.1}$  &-3.46&-0.54\\
Mrk\,590    &$  2.17$& 1.31& 21.26$\pm$0.2 & 11.21$\pm$0.1  &  $1.4^{+0.3}_{-0.3}$ &-4.06&-0.89 \\
Mrk\,817    &$  4.01$& 1.19& 21.17$\pm$0.4 & 11.17$\pm$0.15  &  $
3.7^{+1.1}_{-0.9}$  &-3.61&-1.54\\
NGC\,3227   &$  5.53$& 1.04& 20.46$\pm$0.4 & 10.84$\pm$0.25  &  $4.9^{+2.7}_{-4.9}$ &-3.15 &-1.98\\
NGC\,3783   &$  4.1$& 0.65& 20.07$\pm$0.2 & 10.66$\pm$0.1  &  $1.1^{+1.1}_{-1.0}$  &-3.62&-2.10\\
NGC\,4051   &$ 1.23$& 0.81& 19.70$\pm$0.2 & 10.62$\pm$0.1  &  $0.14^{+ 0.15}_{-0.09}$ &-4.42&-0.86\\
NGC\,4151   &$  5.23$& 0.48& 19.98$\pm$0.4 & 11.04$\pm$0.15  &  $1.2^{+ 0.8}_{-0.7}$ &-3.54 &-2.49\\
NGC\,5548 &$  5.50$& 1.26&  20.89$\pm$0.2& 11.05$\pm$0.1  &  $6.8^{+ 1.5}_{-1.0}$  &-3.06&-1.83\\
NGC\,7469   &$  3.2$& 0.70&20.90$\pm$0.2& 11.05$\pm$0.1 &  $0.76^{+ 0
.75}_{-0.76}$& -4.15& -1.85\\
PG\,0953+414&$ 3.14$& 2.03& 20.29$\pm$1.0\tablenotemark{b}& 11.49$\pm$0.4  &  $15.5^{+ 10.8}_{-9.1}$  &-2.57&-0.49\\
{}&{}&{}&{}&{}&{}&{}&{}\\
\tableline
\end{tabular}
\tablenotetext{a}{Unknown Hubble type, bulge correction estimated assuming Sa}
\tablenotetext{b}{From Bahcall \et (1997)}
\end{center}
\end{table}

We relate the bulge luminosity to the magnitude by the standard expression
$\lg (L_{bulge}/L_\odot )= 0.4 (-M_v + 4.83)$.
The bulge mass is then calculated using the mass-to-light relation for normal
galaxies, 
${M/\ms \over L/L_\odot}~\approx 5 (L/10^{10}L_\odot )^{0.15}$
(see Faber \et 1997).

Fig. 1 shows the \bh~ mass as a function of the bulge mass.
All the objects in our sample have BBRs lower than 0.006,
the average value found for normal galaxies (Magorrian \et 1998, represented
by a dashed line), and the 
sample average is
$<M_{BH}>= 3\times 10^{-4} <M_{bulge}>$.
Also shown is NGC 1068 (a Seyfert 2), with the MBH mass estimated by
maser dynamics.
The narrow-line Seyfert galaxy NGC 4051, which has by far the lowest BBR
in our sample, may indicate that narrow-line Seyfert 1 galaxies have 
smaller \bh s than ordinary Seyfert 1 galaxies
(Wandel and Boller 1998).

\myfig {12} 1 {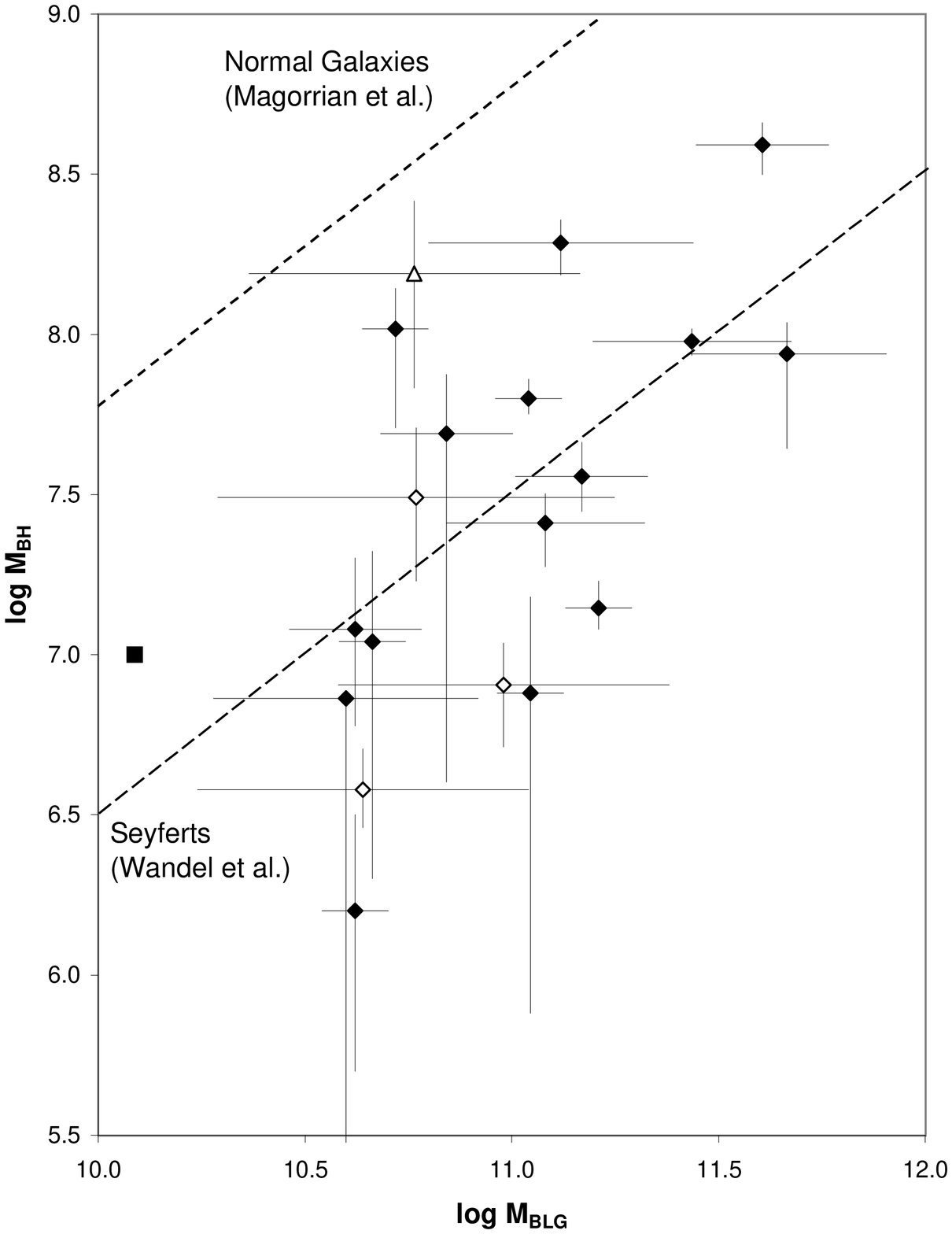}
{1. The virial \bh ~mass calculated by the reverberation BLR
method (from Wandel, Peterson \& Malkan 1999) vs. the bulge magnitude 
(from Whittle 1992) for the 
Seyfert 1 galaxies in our sample (diamond), the masing Seyfert 2 galaxy NGC 1068
(square) and PG0953+414 (triangle). Open diamonds indicate an unknown Hubble 
type (and therefore a large uncertainty in the bulge magnitude).
The dashed diagonal lines are the average BBRs for normal galaxies
(Magorrian \et 1998) and Seyfert 1s (this work).
}

\subsection {Quasars}

Laor (1998) has studied the \bh -host bulge relation for a sample
of 15 bright PG quasars.
Estimating the bulge masses from the Bahcall \et (1997) study of 
quasar host galaxies he admits the uncertainty
in estimating bulge luminosity, dominated by the much 
brighter nonstellar source.

Laor estimates the \bh~ mass using the H$\beta$ line width and the empirical
relation 
$r_{BLR}=15 L_{44}^{1/2}~ ~{\rm light-days}$
 (Kaspi \et 1997),
where $L_{44}=L(0.1-1\micron )$ in units of $10^{44}\ers$.

As this relation has been derived for less than a dozen low- and medium 
luminosity objects (mainly Seyferts) with measured reverberation sizes,
it is not obvious that it may be extrapolated to more luminous quasars.
The BLR size is also dependent on the ionizing and soft X-ray continua
(Wandel 1997).
The WPM sample (which includes Kaspi's sample)
indicates that the slope of the BLR-size
luminosity relation may flatter than 0.5; WPM find
$r\sim 17L_{44}^{0.36\pm 0.09}{\rm l-d}$. 
If this result is correct,
extrapolating the $r\sim L^{1/2}$ relation over two orders of magnitude
(the difference between the average luminosity of the PG quasars used by Laor and Kaspi's sample average) overestimates the \bh~ mass.
Indeed, for the only object common to the Laor and WPM samples -
the quasar PG 0953+414 - Laor finds 3$\times 10^8\ms$, while the \rev -rms method gives $(1.5^{+1.1}_{-0.9})\times 10^8 \ms$.

\subsection {Comparing Normal Galaxies, Seyferts and Quasars}

Fig. 2 shows the three groups in the plane of \bh~mass vs. 
bulge luminosity.
The best fits and the corresponding standard deviations to the 
data in the three groups are
($M_8=M_{BH}/10^8\ms$ and $L_{10}=L_{bulge}/10^{10}L_\odot$):
\begin{enumerate}
\item
Normal galaxies (Magorrian \et 1998, table 2, excluding upper limits) - 
$M_8= 2.9 L_{10}^{1.26}$, $\sigma = 0.47$
\item
PG quasars (Laor 1998, all objects in his table 1) - 
$M_8= 1.6 L_{10}^{1.10}$, $\sigma = 0.38$
\item
Seyfert 1s (this work, excluding NGC 4051)
$M_8= 0.2 L_{10}^{0.83}$, $\sigma = 0.43$
\end{enumerate}
\myfig {13} {1} {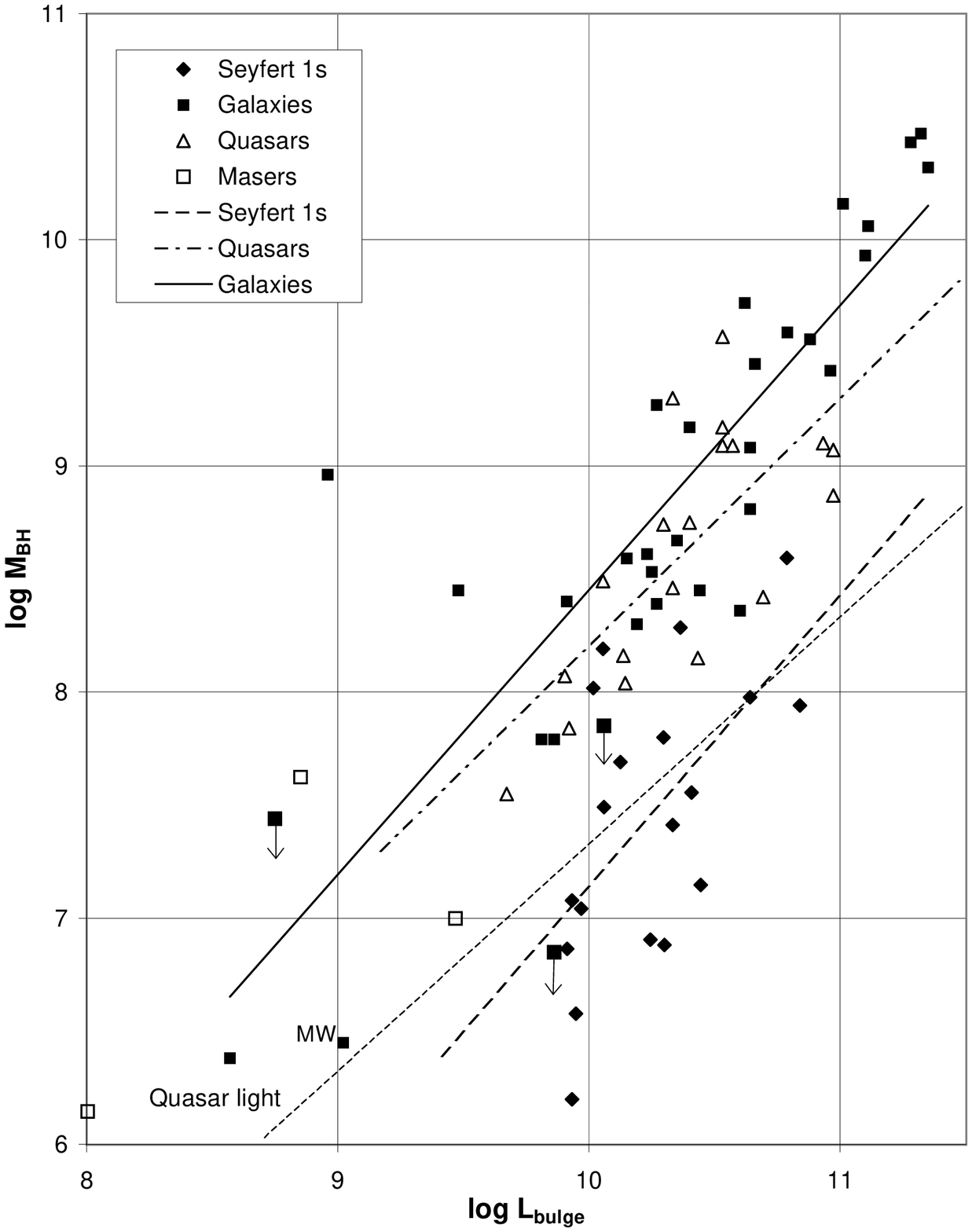}
{2. Mass estimates of MBHs  plotted against the
luminosity of the bulge of the host galaxy.
Squares: MBH candidates from Magorrian \et (1998), 
open squares - MBHs detected by maser dynamics
triangles - 
PG quasars from Laor (1998), diamonds - Seyfert 1 galaxies (this work).
MW denotes our Galaxy.
Also given are the best linear fits for each class (see text).
The dashed long line is the estimate of
dead \bh s from integrated AGN light.
}

As a group Seyfert 1 galaxies have a significantly lower BBR
than normal galaxies and bright quasars.
This lower value agrees with the remnant \bh~ density derived from
integrating the emission from quasars (Chokshi and Turner 1992):
$\rho_{BH}=\int\int (L/\epsilon c^2)\Phi(L,t)dLdt =2\times 10^5(\epsilon /0.1)
\ms {\rm Mpc}^{-1}$,
($\Phi$ is the quasar luminosity function and $\epsilon$ is the efficiency),
which
compared to the density of starlight in galaxies ( $\rho_{gl}$) gives 
  $\rho_{BH}/\rho_{gl}= 2\times 10^{-3} (0.1/\epsilon )(\ms /L_\odot )$
(shown as a dashed line in Fig. 2).

\section {Black hole Evolution and the Black Hole - Bulge ratio}
\subsection{Demography}
 
While Seyfert 1 galaxies seem to have a lower BBR than bright quasars and the galaxies with {\it detected} MBHs in the Magorrian \et (1998) sample, 
they are in good agreement with the BBRs of the {\it upper limits} and of our Galaxy, and with remnant quasar \bhs.
It is plausible therefore that the Seyfert galaxies in our sample represent a larger
population of galaxies with low BBRs, which is under-represented in
the Magorrian \et sample.
This hypothesis is supported by the distribution of \bh~ masses in Fig. 2:
 the only MBHs under 
~$2\times 10^8\ms$ detected by stellar dynamical methods are in the Milky Way, in Andromeda and its satellite M32, and in NGC 3377 (the latter being nearly $10^8\ms $). These galaxies, as well as 
NGC1068 and at least two of the three upper limit in Magorrian's sample
 do have low BBRs, comparable to our Seyfert 1 average.
Actually for angular-resolution limited methods, the MBH detection limit 
is correlated with bulge luminosity: for more luminous
bulges the detection limit is higher, because the stellar velocity
dispersion is higher (the Faber-Jackson relation). In order to detect the 
dynamic effect of a MBH it is necessary to observe closer to the center,
while the most luminous galaxies tend to be at  larger distances, so for
a given angular resolution, the MBH detection limit is higher.

This may imply that Magorrian \et `s sample is biased towards larger
MBHs, as present stellar-dynamical methods are ineffective
for detecting MBHs below $\sim 10^8\ms$ (except in the nearest galaxies).
The BLR method 
is not subject to this constraint,
making Seyfert 1 galaxies good candidates for detecting  
low-mass MBHs.
(Note however that by the same token the WPM sample may be biased towards
Seyferts with low \bh~masses, which tend to vary on shorter timescales
and hence are more likely to be chosen for reverberation studies).

\subsection{Black Hole Growth by Accretion}
Fig. 2 shows that Seyfert 1 galaxies have relatively small MBHs compared 
MBHs in normal galaxies and to quasars, yet they have comparable bulges. 
Below we suggest a possible explanation. 

Consider MBH growth by accretion from the host galaxy.
Since the accretion radius, $R_{acc}\approx 0.3 M_6 v_2^{-2} $pc
(where $M_6=M_{BH}/10^6\ms $ and $v_2=v_*/100\kms$ is the stellar velocity 
dispersion) is small compared with the size of the 
bulge, we may assume the mass supply to the \bh~is given by the 
spherical accretion rate,
$\dot M= 4\pi\lambda R_{acc}{v_*}^2 \rho = (10^{-4}\ms /{\rm yr}) 
\lambda M_6^2v_2^{-3}\rho_*$,
where $\rho_*$ is the stellar (or gas) mass density in units of
$ \ms pc^{-3}$ (corresponding to $4.4 g/cm^3$) and 
$\lambda<1$ is the Bondi parameter combined with a possible reduction factor 
due to angular momentum.
Integrating we find the time required for growing from a mass $M_i$ to $M_f$ 
by accretion of gas or stars,
$t_{acc}= (10^{16} yr) v_2^3\rho_*^{-1}\lambda^{-1} ( \ms / {M_i} - 
 \ms / {M_f} ) \approx (10^8 {\rm yr})M_8^{-1}v_2^3\rho_*^{-1}$.
For masses $<10^6\rho_*^{-1}\ms$ this is larger than the Hubble time, 
so seed MBHs must grow by \bh~
coalescence 
which, even for dense clusters, is of the order of the Hubble time (Lee, 1993; 
Quinlan \& Shapiro 1987). For densities as high as 
in the central parsec of the Milky Way (few$\times 10^7 \ms pc^-3$; Genzel \et 1997) or for NGC 4256 (Miyoshi \et 1995)
accretion-dominated growth becomes feasible
for masses as low as 100-1000$\ms$.  

While the accretion rate is growing as $M^2$ and the growth time decreases as $M^{-1}$, the MBH eventually becomes large enough for accretion-dominated growth
time $t_g=t_{acc}$.
This phase may be applicable for the Seyfert population.
Since the luminosity is  $L\propto\dot M\propto M^2$, the Eddington ratio 
increases as $L/L_{Edd}\propto M\propto L^{1/2}$.
The \bh~growth slows down when the Eddington ratio approaches unity,  
$t_g$ being bound by the Eddington time, 
$
t_g\sim t_E=M/ {\dot M_E} = 4.5\times 10^7  ( \epsilon /{0.1} )^{-1} {\rm yr}
$
where $\dot M_E$ is the accretion rate that would produce an Eddington luminosity.
Equating $t_E$ to $t_{acc}$ we find that the growth rate flattens at
a BH mass of $M_{t}\approx (2\times 10^8\ms ) 
v_2^{3}\rho_*^{-1}(\epsilon/0.1\lambda)$.
In the \ed -limited era, which may correspond to quasars,
the growth rate is exponential, 
depleeting the available matter in the
bulge on the relatively short time scale $t_{Edd}$. This leads to
an asymptotic BBR,
which is likely to be similar for luminous quasars and their largest remnant 
MBHs in normal galaxies.

This scenario predicts that on average quasars should have higher
\ed ratios (near unity) than Seyferts, and larger BBRs.
We can test the prediction from the data at hand.
Estimating the bolometric luminosity of AGN with reverberation data from the lag (WPM), 
and of PG quasars from the relation $L_{bol}\approx 8 \nu 
L_\nu(3000\AA)$ (Laor 1998), 
we find a correlation between the \ed ratio and the BBR,
with Seyferts having \ed ratios in the $10^{-3}-0.1$ range and a low BBR, 
and quasars with Eddington ratios close
to unity and higher BBRs.
From the \ed ratio we can also infer the actual growth time,  
$t_g\approx t_E L/L_{E}$. For most objects in our sample  
$t_g$ is in the range $10^8-{\rm few}\times 10^9$ yr.

\acknowledgments
I acknowledge valuable discussions with Mark Whittle Gary Kriss,
Geremy Goodman, Doug Richstone and Mark Morris
and the hospitality of the Astronomy Department at UCLA.



\begin{references}
Bahcall, J.N., Kirhakos,S.,Saxe,D.H. \& Schneider, D.P. 1997, ApJ 479, 642
\reference {}Blandford, R.D. and McKee, C.F. 1982 Ap.J. 255, 419
\reference{}Chokshi, A. \& Turner, E.L. 1992, MNRAS 259, 421
\reference   {}Faber, S.M. \et 1997, AJ, 114, 1771
\reference {} Blandford, R.D. \& McKee, C.F. 1982, ApJ, 255, 419
\reference{}Genzel R. \et 1997, MNRAS, 264, 455
\reference{}Ghez, A. \et 1998, ApJ 509, 678
\reference{} Kaspi, S. 1997, in Emission Lines in Active Galaxies: New Methods
and Techniques, ed. B.M.Peterson, F.-Z. Cheng and A.S.Wilson 
(San Fransisco: ASP), p. 159
\reference{}
Koratkar, A.P., \& Gaskell, C.M. 1991, ApJS, 75, 719
\reference   {} Kormendy, J., \& Richstone, D.  1995, AR\&A, 33, 581
\reference   {}
Krolik, J.H., Horne, K., Kallman, T.R., Malkan, M.A.,
Edelson, R.A., \& Kriss, G.A. 1991, ApJ, 371, 541
\reference{}Laor, A. 1998, ApJL 505, L83
\reference   {}Lynden-Bell, D. 1969, Nature, 223, 690
\reference   {}Magorrian, J. \et 1998, AJ, 115, 2285.
\reference   {}Miyoshi, M., \et 1995, Nature, 373, 127
\reference   {}Lee, H.L. 1993, ApJ, 418, 147
\reference{}Murray, N. 1998, ApJ 494, 125
\reference {} Nelson, C.H. \& Whittle, M. 1995 ApJS 99 67
\reference {} Nelson, C.H. \& Whittle, M. 1996 ApJ 465, 96
\reference   {}Netzer, H., \& Peterson, B.M. 1997, in Astronomical Time Series,
ed.\ D.\ Maoz, A.\ Sternberg, and E.M.\ Leibowitz, (Dordrecht: Kluwer), 
p.\ 85
\reference   {}Peterson, B. \& Wandel, A 1999, Ap.J.Lett, submitted 
(astro-ph/9905382)
\reference   {}Quinlan, G.D. \& Shapiro, S.L. 1987,  ApJ. 321, 199
\reference {} Rees, M.J. 1984, AR\&A, 22, 471
\reference{}Simien, F. \& de Vaucouleurs, G 1986, ApJ, 302, 564
\reference   {}Richstone, D. \et 1998, Nature 395, A14
\reference   {}Shuder, J. M. 1981, Ap.J. 244, 12
\reference{}Wandel, A. 1997, ApJ, 430, 131
\reference{}Wandel, A. 1999, in Structure and Kinematics of Quasar Broad Line Regions,
ed.\ C.M.\ Gaskell, W.N.\ Brandt, D.\ Dultzin-Hacyan, 
M.\ Dietrich, and M.\ Eracleous (San Francisco: ASP), in press
(astro-ph/9808171)
\reference   {}Wandel, A. \& Boller, Th. 1998, A\&A, 331, 884
\reference   {}
Wandel, A. \& Mushotzky, R.F. 1986, ApJ, 306, L61
\reference   {}Wandel, A., Peterson, B. and Malkan, M 1999, Ap.J. 
in press,  (astro-ph/9905224) (WPM)
\reference{}Whittle, M. \et 1992, ApJS 79, 49
\reference{}Whittle M. 1992, ApJ, 387, 121
\end{references}
\end {document}